\documentclass[12pt]{iopart}
\usepackage{iopams}
\usepackage{graphicx,bm,color}
\begin{document}\title{Emulation of magneto-optic Faraday effect using ultracold atoms}

\author{Zhen Zheng \& Z. D. Wang}
\address{Guangdong-Hong Kong Joint Laboratory of Quantum Matter, Department of Physics, and HKU-UCAS Joint Institute for Theoretical and Computational Physics at Hong Kong,\\The University of Hong Kong, Pokfulam Road, Hong Kong, China}
\ead{zhenzhen.dr@outlook.com \& zwang@hku.hk}

\vspace{10pt}
\begin{abstract}

We propose an arresting scheme for emulating the famous Faraday effect in ultracold atomic gases.
Inspired by the similarities between the light field and bosonic atoms,
we represent the light propagation in medium by the atomic transport in accompany of the laser-atom interaction.
An artificial magneto-optic Faraday effect is readily signaled by the spin imbalance of atoms,
with the setup of laser fields offering a high controllability for quantum manipulation.
The present scheme is really feasible and can be realized with existing experimental techniques of ultracold atoms.
It generalizes the crucial concept of the magneto-optic Faraday effect to ultracold atomic physics, and opens a new way of quantum emulating and exploring the magneto-optic Faraday effect and associated intriguing physics.

\end{abstract}

\vspace{2pc}
\footnote{This is the version of the article before peer review or editing, as submitted by an author to New Journal of Physics. IOP Publishing Ltd is not responsible for any errors or omissions in this version of the manuscript or any version derived from it. The Version of Record is available online at https://doi.org/10.1088/1367-2630/abdce4}

\maketitle

\section{Introduction} \label{sec-introduction}

Emulation of quantum condensed-matter systems using ultracold atoms is an active area in the studies of quantum simulations \cite{Lewenstein2007aip,Georgescu2014rmp,Safronova2018rmp,Schaetz2013njp}.
This is because ultracold atoms can provide a versatile platform with a series of advantages:
(i) The nontrivial interplay or external fields can be designed by the setup of the laser-atom interaction
that can bring novel physics \cite{Dalibard2011rmp,Goldman2014rpp}.
(ii) The high controllability of the synthetic interplay and fields has promising applications such as exploring intriguing phase transitions \cite{Greiner2002nat,Stoferle2004prl,Spielman2007prl} and critical phenomena \cite{Bloch2005nphys,Andersen2002prl}.
(iii) The absence of the disorder effect or impurities makes the well-isolated system ideally clean \cite{Bloch2012nphys},
and thereby facilitates the investigation for unraveling complex phenomena.
Based on these features, a variety of emulations using ultracold atoms have been successfully proposed in a broad range of interesting topics,
for instance the ferromagnetism \cite{Parker2013nphys},
quantum Hall effect \cite{Aidelsburger2013prl,Miyake2013prl,Barberan2015njp,Barbarino2016njp},
atomtronic circuit \cite{Pepino2009prl,Jendrzejewski2014prl} and its hysteresis \cite{Eckel2014nat},
atom transistor \cite{Zhang2015njp},
and optical solenoid associated with magnetic flux \cite{Wang2018prl}.

In condensed-matter physics, the magneto-optic effect is a fundamental but broad concept in the magnetic mediums.
It has been known that the transverse conductivity plays a crucial role in the magneto-optic effect \cite{Pershan2004jap},
which can be introduced by
the interplay of the band exchange splitting and spin-orbit coupling in the magnetic medium \cite{Ebert1996rpp}.
When the light transmits from vacuum to the medium,
the presence of the transverse conductivity hybrids the two polarized components of the photons
and imposes a coherent phase to them during the light propagation.
As the result, the polarized angle of the reflected and forward scattered light fields deviates from the one of the incident light, respectively known as the magneto-optic Kerr effect (MOKE) and Faraday effect (MOFE).
In recent studies, the magneto-optic effect can also arise by virtue of the topological Hall effect,
in which the rotated polarized angle is related to the topological invariant, known as the topological magneto-optic effect \cite{Tse2010prl,Tse2011prb,Feng-2020natcommun}.

However, a great deal of experiment advances on the magneto-optic effect has constituted the focus of major efforts to MOKE rather than MOFE.
This is because the distinct measurement of MOFE has been elusive so far in ordinary magnetic mediums.
In MOKE, the rotation of the polarized angle, which is the prominent feature of the magneto-optic effect,
is affected by the medium boundary condition of the light reflection.
By starkly contrast in MOFE, it accumulates during the light propagation in mediums.
For the sake of the photon absorption by the medium, MOFE is generally expected to be detected in ultra-thin films.
The magnitude of the rotated polarized angle thus dramatically drops in thinner films,
and therefore the salient signal of MOFE is challengingly attainable in real experiments.
On the other hand as mentioned, the emulation using ultracold atoms has advantages in realizing artificial physics system with controllable manipulations.
This motivates us to search a possible alternative routine for studying MOFE via the emulation in the atomic gases, instead of the challenged detection in conventional solid-state systems.

In this work, we propose such a proposal for emulating MOFE using ultracold atoms.
The mechanism for the synthetic MOFE relies on the light-atom interplay,
which stands out from the conventional physics picture
and provides full controllability as well as detectable signals with existing techniques.
The paper is organized as follows.
In section \ref{sec-model}, we present the detailed model of the emulation.
For simplicity we firstly consider the model at resonance to extract the physics picture, and in section \ref{sec-detuned} without loss of generality, the detuned case is investigated.
In section \ref{sec-discussion}, we address the relevant practical considerations and possible implementation of the proposal.
In section \ref{sec-conclusion}, we summarize the work.

\section{Model} \label{sec-model}

We consider the bosons with two internal levels that are denoted as pseudo-spins $\uparrow$ and $\downarrow$.
In ultracold Bose gases, the atomic cloud can be loaded into two reservoirs separated by a mesoscopic channel \cite{Brantut2012sci,Krinner2013prl,Krinner2014nat,Chien2015nphys}.
By preparing the two reservoirs with a number imbalance,
the atomic current can be observed through the channel,
and the hydrodynamics of the atomic cloud density is semi-classically refined by the equation
$\partial_t n + {\bf v}\cdot\nabla n = 0$ \cite{Dalfovo1999rmp}.
The linear dispersion shares the similarity of the light propagation.
Furthermore, in Bose gases, specifically the Bose-Einstein condensate (BEC), the wave function of different pseudo-spins are orthogonal, exhibiting the same property of the polarized components of the light.
Therefore, the atomic transport process inspires us to draw an analogy to the light field in terms of the bosonic cloud.
Despite that the atomic ensemble is totally different from the magnetic mediums, the phenomenal and intrinsic similarities can reveal the fundamental physics at the macroscopic level, which is the focus of quantum simulation.

\begin{figure}[t]
	\centering
	\includegraphics[width=0.8\textwidth]{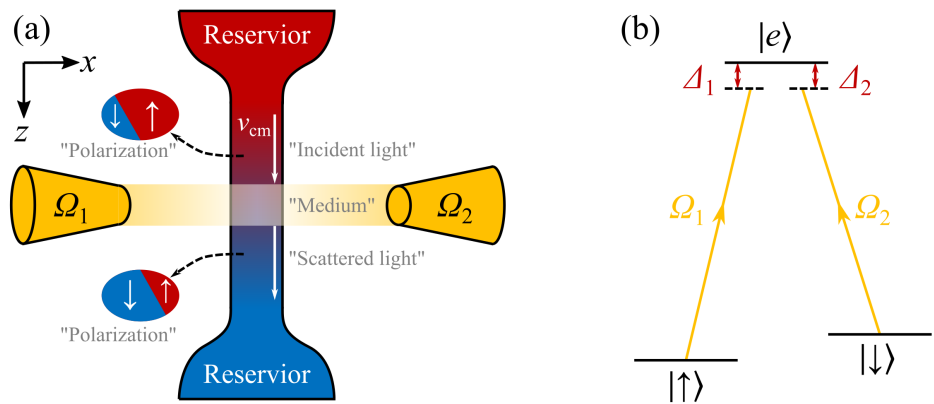}
	\caption{(a) The experimental setup of the proposal: the light field is emulated via the atomic current between two reservoirs, in which the polarization of the light is characterized by the atomic spin imbalance.
	The laser-atom interacting region (yellow) plays the role of the medium in which the light can propagates.
	(b) Illustration of the atomic $\Lambda$-type transition:
	the two pseudospin states of atoms are coupled via a third excited states $|e\rangle$ by means of two Raman lasers $\Omega_{1,2}$ (yellow arrows). The detuning of each laser-atom interaction is denoted as $\Delta_{1,2}$.}
	\label{fig-setup}
\end{figure}

The experimental setup is illustrated in Figure \ref{fig-setup}(a).
We suppose the atomic cloud is prepared in the BEC phase.
In the channel between the reservoirs,
the atomic cloud can be approximately regarded as being confined in the harmonic trap potential $V_{\rm trap}({\bf r})=\frac{1}{2}m\omega_{\rm trap}(x^2+y^2)$.
In the section normal to the $z$ direction, the wave function of the ground state can be given by
\begin{equation}
	\psi({\bf r}) = e^{-(x^2+y^2)/(2l_0^2)}/(\pi l_0^2) \,, \label{eq-inplane-wave}
\end{equation}
where $l_0=\sqrt{\hbar/(m\omega_{\rm trap})}$ \cite{Pethick2008book}.
Along the $z$ direction, the atomic cloud flows at a center-of-mass (COM) velocity $v_{\rm cm}$.

As the spin of atoms mimics the polarization of light,
we implement counter-propagating lasers along the $x$ direction,
which drives a Raman transition between the two spins via an auxiliary excited levels.
In this way, the artificial transverse conductivity can be equivalently generated by laser fields that couples different spins,
and the laser-atom interacting region will play the role of the ``medium''.
The transition is sketched in Figure \ref{fig-setup}(b).
At low temperature, we assume the velocity fluctuation is much smaller than the laser field strength.
In the COM reference frame, such a $\Lambda$ system is governed by the following Hamiltonian,
\begin{equation}
	H = [ \hat{\Omega}_1({\bf r})e^{-i\omega_1t}|e\rangle\langle \uparrow| + \hat{\Omega}_2({\bf r})e^{-i\omega_2t}|e\rangle\langle \downarrow| + H.c. ]
	+ \sum_{\lambda=\uparrow,\downarrow,e}\Gamma_{\lambda}|\lambda\rangle\langle \lambda| \,. \label{eq-h-3level}
\end{equation}
Here $|\lambda\rangle$ with $\lambda=\uparrow,\downarrow,e$ denote the spin-$\uparrow,\downarrow$ and excited states, respectively.
$\Gamma_{\uparrow,\downarrow,e}$ are their corresponding level energies.
$\hat{\Omega}_{\alpha=1,2}({\bf r})\equiv M_{\alpha}({\bf r})e^{ik_\alpha x}$
where $M_{\alpha}({\bf r})$ characterizes the laser field mode followed with
the frequency $\omega_{\alpha}$ as well as the standing-wave vector $k_\alpha$.
$H.c.$ stands for the Hermitian conjugation.
We can assume the general form of the wave function for the three-level system as
$|\psi\rangle = \sum_{\lambda}c_\lambda e^{-i\Gamma_\lambda/\hbar}|\lambda\rangle$.
According to the Schr\"{o}dinger equation $i\partial_t|\psi\rangle = H|\psi\rangle$,
the coefficient $c_\lambda$ satisfies the following equations,
\begin{equation}
	\cases{
		& $i\hbar\partial_t c_\uparrow = \hat{\Omega}_1^*({\bf r}) e^{i\Delta_1 t/\hbar}c_e$ \\
		& $i\hbar\partial_t c_\downarrow = \hat{\Omega}_2^*({\bf r}) e^{i\Delta_2 t/\hbar}c_e$ \\
		& $i\hbar\partial_t c_e = \hat{\Omega}_1({\bf r}) e^{-i\Delta_1 t/\hbar}c_\uparrow + \hat{\Omega}_2({\bf r}) e^{-i\Delta_2 t/\hbar}c_\downarrow$
	}\,.
	\label{eq-pd-c}
\end{equation}
Here we have denoted the detuning as $\Delta_{1} = \Gamma_e - \Gamma_{\uparrow}-\hbar\omega_1$
and $\Delta_{2} = \Gamma_e - \Gamma_{\downarrow}-\hbar\omega_2$.
For simplicity without loss of generality,
we normalize the atomic densities by the total atomic number of the condensate.
Thus the coefficient $c_\lambda$ obeys the constraint $\sum_{\lambda}|c_{\lambda}|^2=1$ due to the number conversation.

In order to give a simple physics picture for our proposal,
we firstly consider the resonance condition $\Delta_1=\Delta_2=0$.
The atoms are initially prepared to reside in the two spin states that host the lowest energy.
Due to spontaneous breaking the $U(1)$ symmetry,
the BEC hosts distinguished phases for each spins \cite{Andrews1997sci}.
Thereby, the initial state of the spinful system can be assumed as the form
$|\psi_0\rangle = \cos\theta\mbox{$|\uparrow\rangle$} + \sin\theta e^{i\varphi}\mbox{$|\downarrow\rangle$}$.
Here $\theta$ characterizes the number imbalance of spins,
and $\varphi$ describes the relative phase between the two spins.
The solution to Eq.(\ref{eq-pd-c}) is then given as follows,
\begin{equation}
	\cases{
		& $c_\uparrow = \cos\theta + F({\bf r})\frac{\hat{\Omega}_1({\bf r})}{\hat{\Omega}_R({\bf r})}\{\cos[\hat{\Omega}_R({\bf r})t/\hbar]-1\}$ \\
		& $c_\downarrow = \sin\theta e^{i\varphi} + F({\bf r})\frac{\hat{\Omega}_2({\bf r})}{\hat{\Omega}_R({\bf r})}\{\cos[\hat{\Omega}_R({\bf r})t/\hbar]-1\}$ \\
		& $c_e = -i F({\bf r})\sin[\hat{\Omega}_R({\bf r})t/\hbar]$
	}\,.
	\label{eq-wavefunc-evol}
\end{equation}
Here the dimensionless function is written as $F({\bf r})=[\hat{\Omega}_1({\bf r})\cos\theta + \hat{\Omega}_2({\bf r})e^{i\varphi}\sin\theta + H.c.]/[2\hat{\Omega}_R({\bf r})]$.
$\hat{\Omega}_R({\bf r})=\sqrt{|\hat{\Omega}_1({\bf r})|^2+|\hat{\Omega}_2({\bf r})|^2}$ is the Rabi frequency.

For simplicity, we postulate the laser modes $M_{1,2}({\bf r})$ to be slowly varied along the $x$ direction in the atomic cloud.
As the laser fields are applied along the $x$ direction,
the atomic transition driven by them involves no momentum transfer in the $z$ direction,
and hence does not affect the atomic transport.
In the laboratory frame, the Rabi frequency can be approximately expanded as $\hat{\Omega}_R({\bf r}_{\rm cm}+{\bf r}') \approx \hat{\Omega}_R({\bf r}_{\rm cm})+{\bf r}'\cdot\nabla\hat{\Omega}_R({\bf r}_{\rm cm})$.
Here the COM coordinate ${\bf r}_{\rm cm}=v_{\rm cm}t\hat{\bf e}_z$ with $\hat{\bf e}_{x,y,z}$ being the unit vector.
Since the laser fields are spatially uniform along the trajectory direction, quantities that depend only on ${\bf r}_{\rm cm}$ can be regarded as constants hereafter.
We denote the gradient of the laser field as $\nabla\hat{\Omega}_R({\bf r}_{\rm cm})\equiv A\hat{\bf e}_x$.

We remark that the gradient potential $A$ can be attainable in practice, for instance by using Gaussian beams whose center deviates from the atomic COM trajectory,
or the tilted potential that is widely applied in the technique of laser-assisted tunneling \cite{Aidelsburger2013prl,Miyake2013prl}.

In a steady transport case,
the atomic current is incompressible along the trajectory direction.
The density per unit of length along the $z$ direction can thus be obtained by $n_{\lambda} = \int |c_{\lambda}\psi({\bf r})|^2 d x d y$, where the spatial distribution $\psi({\bf r})$ has been given in Eq.(\ref{eq-inplane-wave}).
In particular, for spin-$\uparrow$ atoms, it is expressed as
\begin{eqnarray}
	n_\uparrow &= \int \Big\{ [F({\bf r})]^2\frac{|\hat{\Omega}_1|^2}{2\hat{\Omega}_R^2}[\cos(2\hat{\Omega}_Rt/\hbar)-4\cos(\hat{\Omega}_Rt/\hbar)+3] \nonumber\\
	& + F({\bf r})\frac{\hat{\Omega}_1+\hat{\Omega}_1^*}{\hat{\Omega}_R}[\cos(\hat{\Omega}_Rt/\hbar)+1]\cos\theta
	+ \cos^2\theta \Big\} |\psi({\bf r})|^2 d x d y \,. \label{eq-n-up}
\end{eqnarray}
We suppose the spatial scale of the atomic cloud is tremendously larger than the laser wavelengths, i.e. $k_1l_0,k_2l_0\gg1$.
By using the following mathematical relation,
\begin{equation}
	\frac{1}{\sqrt{\pi c}}\int_{-\infty}^{+\infty} \cos(ax+b)e^{-x^2/c}d x = \cos(b)e^{-a^2c/4} \,,
\end{equation}
the rapid spatial modulated terms such as $\cos(k_1x)$ and $\cos(k_2x+\varphi)$ in $F({\bf r})$ of Eq.(\ref{eq-n-up}) will be averaged to exponentially vanish when integrating out the spatial coordinates.
Then we have
\begin{equation}
	n_\uparrow = K_2(t)\Omega_1^2 + K_1(t)\Omega_1^2\cos^2\theta
	+ \cos^2\theta \,,\label{eq-n-evol-up}
\end{equation}
where the time-dependent functions defined as
\begin{eqnarray}
	K_1(t) &= \frac{1}{\Omega_R^2}[ \cos(\Omega_Rt/\hbar)e^{-t^2/\tau_c^2} - 1] \,, \label{eq-k-func-1}\\
	K_2(t) &= \frac{\mathcal{F}}{2\Omega_R^2}[ \cos(2\Omega_Rt/\hbar)e^{-4t^2/\tau_c^2} - 4\cos(\Omega_Rt/\hbar)e^{-t^2/\tau_c^2} + 3] \,, \label{eq-k-func-2}
\end{eqnarray}
and $\mathcal{F}=(\Omega_1^2\cos^2\theta + \Omega_2^2\sin^2\theta)/(2\Omega_R^2)$.
We have denoted $\Omega_{1,2}=M_{1,2}({\bf r}_{\rm cm})$
and $\Omega_{R}=\hat{\Omega}_R({\bf r}_{\rm cm})$.
The decay time $\tau_c$ is defined as
\begin{equation}
	\tau_c = 2\hbar/(Al_0) \,.
\end{equation}
Likewise, the density evolutions of spin-$\downarrow$ and excited-state atoms are obtained as
\begin{eqnarray}
	n_\downarrow &= K_2(t)\Omega_2^2 + K_1(t)\Omega_2^2\sin^2\theta
	+ \sin^2\theta \,, \label{eq-n-evol-down}\\
	n_e &= \frac{\mathcal{F}}{2} [ 1-\cos(2\Omega_Rt/\hbar)e^{-4t^2/\tau_c^2} ] \,. \label{eq-n-evol-e}
\end{eqnarray}

\begin{figure}[t]
	\centering
	\includegraphics[width=0.9\textwidth]{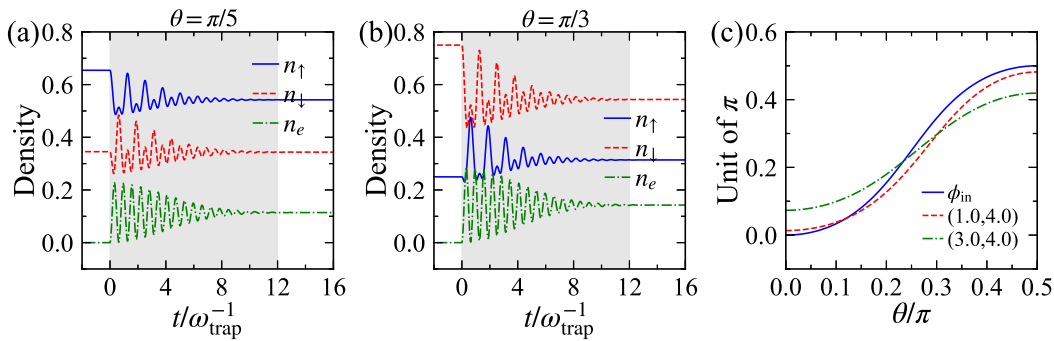}
	\caption{(a)-(b) The evolutions of atomic densities at resonance condition.
	We set $\theta=\pi/5$ in (a) and $\pi/3$ in (b).
	Other parameters are $(\Omega_1,\Omega_2)=(3.0,4.0)\hbar\omega_{\rm trap}$, and $\tau_c^{-1}=0.3\omega_{\rm trap}$.
	The regions of the atomic cloud interacting with lasers are highlighted in gray.
	(c) The polarized angles of the emulated light as functions of $\theta$:
	$\phi_{\rm in}$ (blue-solid line), $\phi_{\rm sc}$ at $(\Omega_1,\Omega_2)=(1.0,4.0)\hbar\omega_{\rm trap}$ (red-dashed line), and $\phi_{\rm sc}$ at $(3.0,4.0)$ (green-dash-dotted line).}
	\label{fig-res1}
\end{figure}

From Eqs.(\ref{eq-n-evol-up}), (\ref{eq-n-evol-down}) and (\ref{eq-n-evol-e}),
one can see that in the presence of the laser field gradient $A$, the Rabi oscillations are exponentially suppressed.
Similar phenomena can be evidenced by experiments yet in a two-level system \cite{Daniel2013pra}.
The atomic cloud will evolve to a steady state in which
the densities of spin $\uparrow$ and $\downarrow$ saturate to
\begin{equation}
	\cases{
		& $n_\uparrow(t\rightarrow \infty) = \cos^2\theta - \frac{\Omega_1^2}{\Omega_R^2}\cos^2\theta + \frac{3\mathcal{F}}{2}\frac{|\Omega_1|^2}{\Omega_R^2}$ \\
		& $n_\downarrow(t\rightarrow \infty) = \sin^2\theta - \frac{\Omega_2^2}{\Omega_R^2}\sin^2\theta + \frac{3\mathcal{F}}{2}\frac{|\Omega_2|^2}{\Omega_R^2}$
	}
	\,.\label{eq-steay-density}
\end{equation}
The dynamic evolutions are shown in Figure \ref{fig-res1}(a) and (b) for different initial setups.
For simplicity, we have assumed the atomic cloud in motion enters the laser region at $t=0$,
and leaves it after the cloud fully evolved to the steady state.
It can be guaranteed by preparing the width of the laser region $L> v_{\rm cm}\tau_c$.

As we use the bosonic atoms to represent the light field,
the polarized angle $\phi$ of the emulated light field is defined by the atomic densities,
\begin{equation}
	\phi = \tan^{-1}[n_\downarrow(t)/n_\uparrow(t)] \,, \label{eq-phi}
\end{equation}
which is time dependent.
In particular, the polarized angle of the incident light is expressed as $\phi_{\rm in}=\tan^{-1}(\tan^2\theta)$,
while for the scattered light is calculated by $\phi_{\rm sc}=\tan^{-1}[n_\downarrow(\infty)/n_\uparrow(\infty)]$ (c.f. Eq.(\ref{eq-steay-density})).
In Figure \ref{fig-res1}(c), we can see the polarized angle is changed after the light passes through the emulated medium, exhibiting the manifest feature of MOFE.
The signal of the artificial MOFE (i.e. the rotated polarized angle) not only depends on the parameter $\theta$ of initial setups, but is also controllable by the laser field strengths $\Omega_{1,2}$.

\section{Detuned case} \label{sec-detuned}

The resonance condition used in the above discussions will introduce the additional heating effect that are frustrated to the practical experiments such as suppressing the lifetime of ultracold atoms \cite{Ketterle1999colle}.
However, we remark that the resonance condition in the $\Lambda$ system is not necessary,
instead the proposal still works when the laser-atom interaction is prepared with a detuning $\Delta_1=\Delta_2=\Delta\neq0$.
It has a crucial advantage that,
at the fully far detuned regime (i.e. $\Delta\gg |\hat{\Omega}_{1,2}({\bf r})|$), the heating effect can prominently suppressed and thereby facilitates the realization of the proposal.

\begin{figure}[t]
	\centering
	\includegraphics[width=0.9\textwidth]{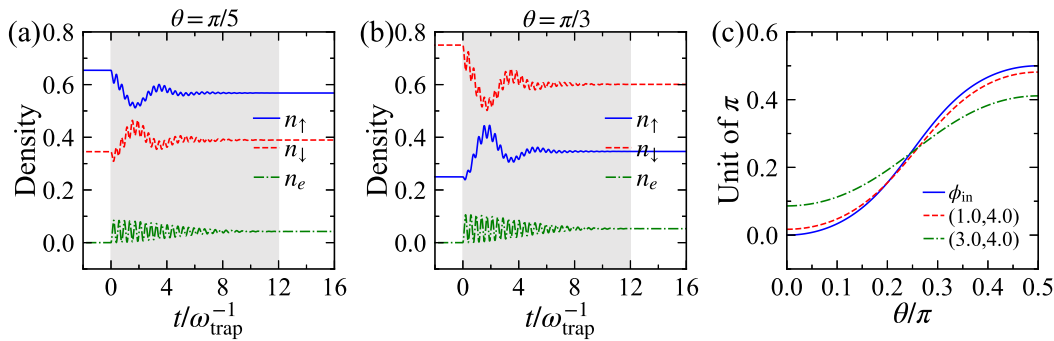}
	\caption{(a)-(b) The evolutions of atomic densities with a detuning $\Delta=13.0\hbar\omega_{\rm trap}$.
	We set $\theta=\pi/5$ in (a) and $\pi/3$ in (b).
	Other parameters are $(\Omega_1,\Omega_2)=(3.0,4.0)\hbar\omega_{\rm trap}$, and $\tau_{c\pm}^{-1}=0.3\omega_{\rm trap}$.
	The regions of the atomic cloud interacting with lasers are highlighted in gray.
	(c) The polarized angles of the emulated light as functions of $\theta$:
	$\phi_{\rm in}$ (blue-solid line), $\phi_{\rm sc}$ at $(\Omega_1,\Omega_2)=(1.0,4.0)\hbar\omega_{\rm trap}$ (red-dashed line), and $\phi_{\rm sc}$ at $(3.0,4.0)$ (green-dash-dotted line).}
	\label{fig-res2}
\end{figure}

At the detuned case, the evolutions of the atomic densities for spin $\uparrow$ and $\downarrow$ share the same forms of Eqs.(\ref{eq-n-evol-up}) and (\ref{eq-n-evol-down}),
but the time-dependent functions are instead rewritten as (see \ref{app-detuned-case})
\begin{eqnarray}
	K_1(t) &= \frac{1}{\Omega_R^++\Omega_R^-}\sum_{\alpha=\pm}\frac{\cos(\Omega_R^\alpha t/\hbar)e^{-t^2/\tau_{c\alpha}^2}-1}{\Omega_R^\alpha} \,, \\
	K_2(t) &= \sum_{\alpha=\pm}\frac{2\mathcal{F}'}{|\Omega_R^\alpha|^2}
	+\frac{2\mathcal{F}'}{\Omega_R^+\Omega_R^-}\cos[(\Omega_R^++\Omega_R^-)t/\hbar]e^{-4t^2/\widetilde{\tau}_c^2} \nonumber\\
	&-\sum_{\alpha,\alpha'}\frac{2\mathcal{F}'}{\Omega_R^{\alpha}\Omega_R^{\alpha'}}
	\cos(\Omega_R^\alpha t/\hbar)e^{-t^2/\tau_{c\alpha}^2}
	+ \frac{2\mathcal{F}'}{\Omega_R^{+}\Omega_R^{-}} \,.
\end{eqnarray}
Here the Rabi oscillations are split into two branches whose frequencies are expresses as
$\hat{\Omega}_R^{\pm}({\bf r})=\Delta/2 \pm \sqrt{|\hat{\Omega}_1({\bf r})|^2+|\hat{\Omega}_2({\bf r})|^2+\Delta^2/4}$.
The dimensionless constant $\mathcal{F}' = (\Omega_{1}^2\cos^2\theta + \Omega_{2}^2\sin^2\theta)/[2(\Omega_R^++\Omega_R^-)^2]$.
We have denoted
$\Omega_R^\pm = \pm\hat{\Omega}_R^\pm({\bf r}_{\rm cm})$, $A_\pm = \pm\nabla_x \hat{\Omega}_R^\pm({\bf r}_{\rm cm})$, $\tau_{c\pm} = 2\hbar/(A_\pm l_0)$, and $\widetilde{\tau}_c^{-1} = \tau_{c+}^{-1} + \tau_{c-}^{-1}$.
It is easy to demonstrated that $K_{1,2}(t)$ reduces to the form given in Eqs.(\ref{eq-k-func-1}) and (\ref{eq-k-func-2}) at resonance $\Delta=0$.
The evolutions are plotted in Figure \ref{fig-res2}(a) and (b).
Likewise as in a similar way to the resonance condition,
the polarized angle of the incident light is shifted after passing through the medium, as shown in Figure \ref{fig-res2}(c).

We comment that the decay time $\tau_{c+}$ and $\tau_{c-}$ are indeed identical. This is because the spatial dependence of $\hat{\Omega}_{R}^{\pm}({\bf r})$ originates from the laser field modes $M_{1,2}({\bf r})$, and hence their gradients $A_\pm$ are equal to each other.
In comparison between Figure \ref{fig-res1}(c) and Figure \ref{fig-res2}(c),
we find that the rotated polarized angles is insensitive to the detuning $\Delta$.
This is because in the steady state, $\Delta$ only affects the Rabi frequencies $\hat{\Omega}_{R}^{\pm}({\bf r})$,
which is nearly canceled out in the calculation using Eq.(\ref{eq-phi}).

\section{Discussions} \label{sec-discussion}

\subsection{Non-condensed components}
In general, the Bose gas is composed of not only the BEC component but also the non-condensed one.
This is is usually evidenced by the contrast of condensate densities in spatial and momentum spaces \cite{Anderson1995sci,Davis1995prl,Bradley1995prl}.
In the plane normal to the trajectory direction, the spatial profile of the non-condensed wave function can be given by
$\psi_{\rm nc}({\bf r})=e^{-(x^2+y^2)/(2l_{\rm nc}^2)}/(\pi l_{\rm nc}^2)$ with $l_{\rm nc}=\sqrt{k_BT/(m\omega_{\rm trap}^2)}$ at temperature $T$.
Similar to the results of the BEC case, the laser field gradients can also lead to the exponential damping in the evolution of atomic densities.
However, the decay times are approximately estimated as $\tau_{c\pm}^{\rm nc}=\tau_{c\pm}l_0/l_{\rm nc}$ instead.
It implies that the densities of the non-condensed component will evolve to the steady state faster at higher temperature.

The presence of the non-condensed component does not alter the results or arguments obtained before.
This can be explained as follows.
Outside the laser-atom interacting region, there is no coupling between the two spins.
The system of each spin reduces to the ideal Bose gas confined in a two-dimensional harmonic trap potential.
It can be demonstrated that (see \ref{app-tube-BEC}), in a bosonic system composed of $N_{\rm total}$ atoms, the atomic number of the non-condensed component $N_{\rm nc}$ is determined by $N_{\rm nc}/N_{\rm total}=\mbox{$(T/T_c)^{5/2}$}$ below $T_c$.
Here $T_c$ stands for the critical temperature for the phase transition that BEC vanishes.
We can find that $N_{\rm nc}$ depends only on the temperature $T$, and is proportional to the BEC number:
$N_{0}=N_{\rm total}-N_{\rm nc}=[\mbox{$(T_c/T)^{5/2}$}-1]N_{\rm nc}$.
Therefore, the density ratio between opposite spins in the non-condensed component is identical to the results obtained in BEC.
It reveals the polarized angles of the two components evolve in the same way, yet are damped in different decay time.

\subsection{Interaction effect}
The theoretical results obtained in the above sections are based on the single-particle properties.
By choosing proper atom samples, the intrinsic inter-atomic interaction originated from the van der Waals potential can be weak and ignorable in the transport process.
For example, it is known that the scattering length of $^{88}$Sr atoms is approximately $-2a_0$
with $a_0$ being the Bohr radius \cite{deEscobar2008pra,Stellmer2009prl}.
The use of $^{88}$Sr can decrease the interaction effect close to zero,
thereby the obtained results will maintain the accuracy.
On the other hand, we remark that the principal idea of an MOFE emulator in this proposal works as long as the spin-imbalanced densities of the final state explicitly depends on the initial setup.
Under weak interaction like the contact one,
the atom number and the spin imbalance are both conservative.
For this sake, the emulation and observation of MOFE obtained by atomic densities will be still valid when the single-particle properties dominate the physics of the system.

\subsection{Spontaneous emission effect}
The excited state $|e\rangle$ is occupied even after the atoms evolve to the steady state (c.f. Eq.(\ref{eq-wavefunc-evol})).
As the atomic density of each spin respectively characterizes the amplitudes of the polarized light field,
the residence number $n_e$ can be used to represent the absorbance ratio of the the emulated medium.
However, the ubiquitous spontaneous emission of the atoms will lead the decay from the excited state $|e\rangle$ to the two spin states that host the lower energy.
In the $\Lambda$ system of the proposal,
there are two dressed states that are mutually orthogonal: the bright one $|\psi_{B}\rangle = \Omega_1/\Omega_R\mbox{$|\uparrow\rangle$} + \Omega_2/\Omega_R\mbox{$|\downarrow\rangle$}$ which is coupled to $|e\rangle$ through the laser fields, and the dark one $|\psi_D\rangle = -\Omega_2/\Omega_R\mbox{$|\uparrow\rangle$} + \Omega_1/\Omega_R\mbox{$|\downarrow\rangle$}$ which is decoupled from $|e\rangle$ and $|\psi_B\rangle$.
For the sake of the spontaneous emission, the atoms eventually evolve to the dark state $|\psi_D\rangle$.
The dynamic evolution is known as the coherent population trapping \cite{Scully1997book} and is widely applied in the laser cooling \cite{Aspect1988prl,Aspect1989josab}.
The spin imbalance of the dark state, i.e. the polarized angle of the scattered light,
is solely determined by the laser field strength $\Omega_{1,2}$ and is independent from the polarized angle of the incident light.
At this time, the laser-atom region plays the role of a polarizer.
It filters the light with a specific polarized angle $\phi=\tan^{-1}(\Omega_1^2/\Omega_2^2)$.
Therefore, in order to realize an emulator of MOFE,
the spontaneous emission effect needs to be suppressed.
In practice, it can be achieved by decreasing the population occupied in $|e\rangle$.
By comparing $n_e$ illustrated in Figures \ref{fig-res1} and \ref{fig-res2},
one can easily find that $n_e$ is nearly empty under the far detuning condition,
and thus the suppression on the spontaneous emission effect is anticipated in this case.
We note that besides the $\Lambda$-type transition, the use of a single optical field that directly couples pseudo-spins may also work for the MOFE emulator, but provides less controllability.

\subsection{Experimental implements}

The proposal is readily realized via current techniques using ultracold atoms.
Here we use $^{87}$Rb as the example to estimate the feasibility of the proposal.
We choose two hyperfine states of $^5$S$_{1/2}$ as pseudo-spins.
The interaction effect can be suppressed by tuning Feshbach resonances.
By setting $\omega_{\rm trap}\approx 2\pi \times 200$Hz,
the condensate lengths are evaluated as $l_0\approx 0.76\mbox{$\mu$m}$ and $l_{\rm nc}\approx 2.4\mbox{$\mu$m}$ at temperature 100nK.
If we choose the laser field gradient as $A\approx 200$kHz/mm,
the decay time is obtained as $\tau_c\approx 2.1$ms and $\tau_c^{\rm nc}\approx 0.65$ms.
Noticing that the Rabi oscillations decay exponentially as $e^{-t^2/\tau_c^2}$,
the system will evolve to the steady state within shorter time than milliseconds.
The Faraday rotation illustrated in Figure \ref{fig-res1}(c) and \ref{fig-res2}(c) can be detected
by preparing the laser fields $\Omega_{1,2}$ to the order of $\hbar\omega_{\rm trap}$.

\section{Conclusions} \label{sec-conclusion}

In summary, we propose a scheme for quantum emulating MOFE that is frustrated to be evidenced in practical experiments.
The core of the quantum simulation relies on the analogy between the classic light field and the bosonic atomic cloud.
Our proposal broadens a classic concept of MOFE to ultracold atomic physics,
and provides an alternative perspective of understanding the laser-atom interaction in atomic ensembles at a macroscopic level.
The feasible measurement with the high controllability and distinguished signals facilitates the observation to the artificial MOFE, and unambiguously paves the way for the quantum emulating and exploring MOFE.

The present work has focused on the physics of the homogeneous laser-atom interaction.
If we design a spatially-dependent or spin-dependent structure for such an interplay,
it is known to support a series of effective external fields that associate with rich physics \cite{Dalibard2011rmp,Goldman2014rpp}.
For instance, based on existing techniques \cite{Lin2016jpb},
it is possible to associate the rotated polarized angle with a nonlinear dependence on external artificial fields,
i.e. the emulation of the magneto-optic Voigt effect.
Another potential application of the present work is to relate the artificial transverse conductivity to the intrinsic topological properties of system,
revealing the possibility of a quantized Faraday rotation.
These represent an interesting line of future research.

\section*{Acknowledgements}

We acknowledge S. Z. Zhang for helpful discussions.
This work was supported by
the Key-Area Research and Development Program of GuangDong Province (Grant No. 2019B030330001),
the National Key Research and Development Program of China (Grant No. 2016YFA0301800),
the GRF (No.: HKU173057/17P) and CRF (No.: C6005-17G) of Hong Kong.

\appendix
\section{Detailed formulas of the detuned case} \label{app-detuned-case}

Here we show the dynamic evolutions at the detuned condition $\Delta_1=\Delta_2=\Delta\neq0$.
By denoting $\hat{c}_e=e^{i\Delta t}c_e$,
the solutions of Eq.(\ref{eq-pd-c}) are expressed as follows,
\begin{equation}
	\cases{
		& $c_\uparrow = F'\hat{\Omega}_1^*\Big( \frac{e^{i\hat{\Omega}_R^+ t/\hbar}-1}{\hat{\Omega}_R^+}-\frac{e^{i\hat{\Omega}_R^- t/\hbar}-1}{\hat{\Omega}_R^-} \Big)+\cos\theta$ \\
		& $c_\downarrow = F'\hat{\Omega}_2^*\Big( \frac{e^{i\hat{\Omega}_R^+ t/\hbar}-1}{\hat{\Omega}_R^+}-\frac{e^{i\hat{\Omega}_R^- t/\hbar}-1}{\hat{\Omega}_R^-} \Big)+\sin\theta e^{i\varphi}$ \\
		& $\hat{c}_e = - F'( e^{i\hat{\Omega}_R^+ t/\hbar}-e^{i\hat{\Omega}_R^- t/\hbar})$
	} \,,
\end{equation}
where $F'({\bf r})$ is given as
\begin{equation}
	F'({\bf r}) = \frac{1}{2(\hat{\Omega}_R^+-\hat{\Omega}_R^-)} \{[\hat{\Omega}_1({\bf r})+\hat{\Omega}_1^*({\bf r})]\cos\theta + [\hat{\Omega}_2({\bf r})e^{i\varphi}+\hat{\Omega}_2^*({\bf r})e^{-i\varphi}]\sin\theta \} \,.
\end{equation}
We can see that the periodic oscillation will be split into two branches
associated with the Rabi frequencies given by
\begin{eqnarray}
	\hat{\Omega}_R^\pm({\bf r})&= \sqrt{|\hat{\Omega}_1({\bf r})|^2+|\hat{\Omega}_2({\bf r})|^2+\frac{\Delta^2}{4}} \pm \frac{\Delta}{2} \\
	&= \sqrt{[M_1({\bf r})]^2+[M_2({\bf r})]^2+\frac{\Delta^2}{4}} \pm \frac{\Delta}{2} \,.
\end{eqnarray}
Specifically at far detuning $\Delta\gg\hat{\Omega}_{1,2}$, the Rabi frequencies reduce to
\begin{equation}
	\hat{\Omega}_R^+({\bf r}) \approx \Delta + \frac{[M_1({\bf r})]^2+[M_2({\bf r})]^2}{\Delta} \,,\quad
	\hat{\Omega}_R^-({\bf r}) \approx \frac{[M_1({\bf r})]^2+[M_2({\bf r})]^2}{\Delta}
	\,. \label{eq-raman-rabi-freq}
\end{equation}

After going through the same approach in obtaining the damping introduced by the laser field gradient,
the density evolutions are written as
\begin{eqnarray}
	n_\uparrow &= K_2(t)\Omega_1^2
	+ K_1(t)\Omega_1^2 \cos^2\theta + \cos^2\theta \,,\\
	n_\downarrow &= K_2(t)\Omega_2^2
	+ K_1(t)\Omega_2^2 \sin^2\theta + \sin^2\theta \,,\\
	n_e &= 2\mathcal{F}'\{ 1-\cos[(\Omega_R^++\Omega_R^-)t/\hbar]e^{-4t^2/\widetilde{\tau}_c^2} \} \,,
\end{eqnarray}
where
\begin{eqnarray}
	& K_1(t) = \frac{1}{\Omega_R^++\Omega_R^-}\sum_{\alpha=\pm}\frac{\cos(\Omega_R^\alpha t/\hbar)e^{-t^2/\tau_{c\alpha}^2}-1}{\Omega_R^\alpha} \,,\\
	& K_2(t) = \mathcal{F}'\sum_{\alpha=\pm}\frac{2-2\cos(\Omega_R^\alpha t/\hbar)e^{-t^2/\tau_{c\alpha}^2}}{|\Omega_R^\alpha|^2} \nonumber \\
	&+\frac{2\mathcal{F}'}{\Omega_R^+\Omega_R^-}\{ \cos[(\Omega_R^++\Omega_R^-)t/\hbar]e^{-4t^2/\widetilde{\tau}_c^2} - \sum_{\alpha=\pm}\cos(\Omega_R^\alpha t/\hbar)e^{-t^2/\tau_{c\alpha}^2} + 1\} \,,
\end{eqnarray}
and
\begin{eqnarray}
	&\Omega_{1,2}=M_{1,2}({\bf r}_{\rm cm}) \,,\quad
	\mathcal{F}' = \frac{\Omega_{1}^2\cos^2\theta + \Omega_{2}^2\sin^2\theta}{2(\Omega_R^++\Omega_R^-)^2} \,,\\
	&\Omega_R^\pm = \pm\hat{\Omega}_R^\pm({\bf r}_{\rm cm}) \,,\quad
	A_\pm = \pm\nabla_x \hat{\Omega}_R^\pm({\bf r}_{\rm cm}) \,,\\
	&\tau_{c\pm} = 2\hbar/(A_\pm l_0) \,,\quad
	\widetilde{\tau}_c^{-1} = \tau_{c+}^{-1} + \tau_{c-}^{-1} \,.
\end{eqnarray}
From Eq.(\ref{eq-raman-rabi-freq}) one can find $\tau_{c+}$ and $\tau_{c-}$ are identical.
At the steady state, we have
\begin{equation}
	\cases{
		& $n_\uparrow(t\rightarrow \infty) = 2\mathcal{F}'\Omega_1^2 \mathcal{D}
		+(1- \frac{\Omega_1^2}{\Omega_R^+\Omega_R^-})\cos^2\theta$ \\
		& $n_\downarrow(t\rightarrow \infty) = 2\mathcal{F}'\Omega_2^2 \mathcal{D}
		+(1- \frac{\Omega_2^2}{\Omega_R^+\Omega_R^-})\sin^2\theta$ \\
		& $n_e(t\rightarrow \infty) = 2\mathcal{F}'$
	}\,,
\end{equation}
where $\mathcal{D}=1/|\Omega_R^+|^2+1/|\Omega_R^-|^2+1/(\Omega_R^+\Omega_R^-)$.

\section{Bose gas in a tube profile}\label{app-tube-BEC}

We consider the Bose gas that is confined in a two-dimensional harmonic trap in the $x$-$y$ plane and is free along the $z$ direction, i.e. a tube configuration in the spatial space.
The dispersion of the system is
\begin{equation}
	E(\epsilon,k_z) = \frac{\hbar^2k_z^2}{2m} + \epsilon \,,\qquad
	\epsilon = (n_x+\frac{1}{2})\hbar\omega_{\rm trap} + (n_y+\frac{1}{2})\hbar\omega_{\rm trap} \,,
\end{equation}
where $n_{x,y}$ stands for the discrete levels in the $x$-$y$ plane.
The total number of states below the energy $\epsilon$ is
\begin{equation}
	G(\epsilon) = \frac{1}{\hbar^2\omega_{\rm trap}^2} \int_0^\epsilon d\epsilon_x \int_0^{\epsilon-\epsilon_x} d\epsilon_y
	=\frac{\epsilon^2}{2\hbar^2\omega_{\rm trap}^2} \,.
\end{equation}
The corresponding density of states is
\begin{equation}
	g(\epsilon) = \frac{d G(\epsilon)}{d \epsilon} = \frac{\epsilon}{\hbar^2\omega_{\rm trap}^2} \,.
\end{equation}
The number of the non-condensed component with a length of $L$ along the $z$ direction is calculated by
\begin{equation}
	N_{\rm nc} = \int_{-\infty}^{+\infty}\int_0^{\infty} \frac{1}{e^{E(k_z,\epsilon)/k_BT}-1} \times
	\frac{\epsilon}{\hbar^2\omega_{\rm trap}^2}d \epsilon \times \frac{L}{2\pi}d k_z \,. \label{eq-n-nc}
\end{equation}
By making variable substitutions as follows
\begin{equation}
	\epsilon=\hbar^2k_\parallel^2/(2m)\qquad \mathrm{and}\qquad(k_z,k_\parallel) = (k\cos\Phi,k\sin\Phi) \,,
\end{equation}
Eq.(\ref{eq-n-nc}) can be calculated as
\begin{equation}
	N_{\rm nc} = \frac{L\hbar^2}{2\pi m^2\omega_{\rm trap}^2}
	\int_0^{\pi/2}\int_0^{\infty}
	\frac{k^4\sin^3\Phi}{e^{\hbar^2k^2/(2m k_BT)}-1} d k d \Phi \,.
\end{equation}
Using the following relation ($\zeta(s)$ is the Riemann zeta function and $\Gamma(s)$ is the Gamma function)
\begin{equation}
	\zeta(s) = \frac{1}{\Gamma(s)}\int_0^\infty \frac{x^{s-1}}{e^x-1}d x \,,
\end{equation}
we obtain
\begin{equation}
	N_{\rm nc} = C_{5/2}(k_BT)^{5/2} \,,\qquad
	C_{5/2} = \frac{L}{\hbar^3\omega_{\rm trap}^2} \sqrt{\frac{m}{2\pi}} \zeta(5/2) \,.
\end{equation}
At critical temperature $T_c$ of the BEC transition, $N_{\rm nc}$ equals to the total number $N_{\rm total}$ of bosons,
yielding
\begin{equation}
	N_{\rm total} = C_{5/2}(k_BT_c)^{5/2} \,.
\end{equation}
Therefore, the number of the BEC component is
\cite{Pethick2008book}
\begin{equation}
	N_0 = N_{\rm total} - N_{\rm nc}= N_{\rm nc}\Big[\Big(\frac{T_c}{T}\Big)^{5/2}-1\Big] \,.
\end{equation}

\section*{References}
\bibliographystyle{iopart-num}
\bibliography{bib}

\providecommand{\newblock}{}
\begin{thebibliography}{10}
\expandafter\ifx\csname url\endcsname\relax
  \def\url#1{{\tt #1}}\fi
\expandafter\ifx\csname urlprefix\endcsname\relax\def\urlprefix{URL }\fi
\providecommand{\eprint}[2][]{\url{#2}}

\bibitem{Lewenstein2007aip}
Lewenstein M, Sanpera A, Ahufinger V, Damski B, Sen(De) A and Sen U 2007 {\em
  Adv. Phys.\/} {\bf 56} 243--379

\bibitem{Georgescu2014rmp}
Georgescu I~M, Ashhab S and Nori F 2014 {\em Rev. Mod. Phys.\/} {\bf 86}
  153--185

\bibitem{Safronova2018rmp}
Safronova M~S, Budker D, DeMille D, Kimball D~F~J, Derevianko A and Clark C~W
  2018 {\em Rev. Mod. Phys.\/} {\bf 90} 025008

\bibitem{Schaetz2013njp}
Schaetz T, Monroe C~R and Esslinger T 2013 {\em New J. Phys.\/} {\bf 15} 085009

\bibitem{Dalibard2011rmp}
Dalibard J, Gerbier F, Juzeli{\ifmmode\bar{u}\else\={u}\fi}nas G and
  {\ifmmode\ddot{O}\else\"{O}\fi}hberg P 2011 {\em Rev. Mod. Phys.\/} {\bf 83}
  1523--1543

\bibitem{Goldman2014rpp}
Goldman N, Juzeli{\ifmmode\bar{u}\else\={u}\fi}nas G,
  {\ifmmode\ddot{O}\else\"{O}\fi}hberg P and Spielman I~B 2014 {\em Rep. Prog.
  Phys.\/} {\bf 77} 126401

\bibitem{Greiner2002nat}
Greiner M, Mandel O, Esslinger T, H{\ifmmode\ddot{a}\else\"{a}\fi}nsch T~W and
  Bloch I 2002 {\em Nature\/} {\bf 415} 39--44

\bibitem{Stoferle2004prl}
St{\ifmmode\ddot{o}\else\"{o}\fi}ferle T, Moritz H, Schori C,
  K{\ifmmode\ddot{o}\else\"{o}\fi}hl M and Esslinger T 2004 {\em Phys. Rev.
  Lett.\/} {\bf 92} 130403

\bibitem{Spielman2007prl}
Spielman I~B, Phillips W~D and Porto J~V 2007 {\em Phys. Rev. Lett.\/} {\bf 98}
  080404

\bibitem{Bloch2005nphys}
Bloch I 2005 {\em Nat. Phys.\/} {\bf 1} 23--30

\bibitem{Andersen2002prl}
Andersen J~O, Al~Khawaja U and Stoof H~T~C 2002 {\em Phys. Rev. Lett.\/} {\bf
  88} 070407

\bibitem{Bloch2012nphys}
Bloch I, Dalibard J and Nascimb{\ifmmode\grave{e}\else\`{e}\fi}ne S 2012 {\em
  Nat. Phys.\/} {\bf 8} 267--276

\bibitem{Parker2013nphys}
Parker C~V, Ha L~C and Chin C 2013 {\em Nat. Phys.\/} {\bf 9} 769--774

\bibitem{Aidelsburger2013prl}
Aidelsburger M, Atala M, Lohse M, Barreiro J~T, Paredes B and Bloch I 2013 {\em
  Phys. Rev. Lett.\/} {\bf 111} 185301

\bibitem{Miyake2013prl}
Miyake H, Siviloglou G~A, Kennedy C~J, Burton W~C and Ketterle W 2013 {\em
  Phys. Rev. Lett.\/} {\bf 111} 185302

\bibitem{Barberan2015njp}
Barber{\ifmmode\acute{a}\else\'{a}\fi}n N, Dagnino D,
  Garc{\ifmmode\acute{\imath}\else\'{\i}\fi}a-March M~A, Trombettoni A, Taron J
  and Lewenstein M 2015 {\em New J. Phys.\/} {\bf 17} 125009

\bibitem{Barbarino2016njp}
Barbarino S, Taddia L, Rossini D, Mazza L and Fazio R 2016 {\em New J. Phys.\/}
  {\bf 18} 035010

\bibitem{Pepino2009prl}
Pepino R~A, Cooper J, Anderson D~Z and Holland M~J 2009 {\em Phys. Rev.
  Lett.\/} {\bf 103} 140405

\bibitem{Jendrzejewski2014prl}
Jendrzejewski F, Eckel S, Murray N, Lanier C, Edwards M, Lobb C~J and Campbell
  G~K 2014 {\em Phys. Rev. Lett.\/} {\bf 113} 045305

\bibitem{Eckel2014nat}
Eckel S, Lee J~G, Jendrzejewski F, Murray N, Clark C~W, Lobb C~J, Phillips W~D,
  Edwards M and Campbell G~K 2014 {\em Nature\/} {\bf 506} 200--203

\bibitem{Zhang2015njp}
Zhang Z, Dunjko V and Olshanii M 2015 {\em New J. Phys.\/} {\bf 17} 125008

\bibitem{Wang2018prl}
Wang B, {\ifmmode\ddot{U}\else\"{U}\fi}nal F~N and Eckardt A 2018 {\em Phys.
  Rev. Lett.\/} {\bf 120} 243602

\bibitem{Pershan2004jap}
Pershan P~S 2004 {\em J. Appl. Phys.\/} {\bf 38} 1482

\bibitem{Ebert1996rpp}
Ebert H 1996 {\em Rep. Prog. Phys.\/} {\bf 59} 1665--1735

\bibitem{Tse2010prl}
Tse W~K and MacDonald A~H 2010 {\em Phys. Rev. Lett.\/} {\bf 105} 057401

\bibitem{Tse2011prb}
Tse W~K and MacDonald A~H 2011 {\em Phys. Rev. B\/} {\bf 84} 205327

\bibitem{Feng-2020natcommun}
Feng W, Hanke J~P, Zhou X, Guo G~Y, Bl{\ifmmode\ddot{u}\else\"{u}\fi}gel S,
  Mokrousov Y and Yao Y 2020 {\em Nat. Commun.\/} {\bf 11} 118

\bibitem{Brantut2012sci}
Brantut J~P, Meineke J, Stadler D, Krinner S and Esslinger T 2012 {\em
  Science\/} {\bf 337} 1069--1071

\bibitem{Krinner2013prl}
Krinner S, Stadler D, Meineke J, Brantut J~P and Esslinger T 2013 {\em Phys.
  Rev. Lett.\/} {\bf 110} 100601

\bibitem{Krinner2014nat}
Krinner S, Stadler D, Husmann D, Brantut J~P and Esslinger T 2014 {\em
  Nature\/} {\bf 517} 64--67

\bibitem{Chien2015nphys}
Chien C~C, Peotta S and Di~Ventra M 2015 {\em Nat. Phys.\/} {\bf 11} 998--1004

\bibitem{Dalfovo1999rmp}
Dalfovo F, Giorgini S, Pitaevskii L~P and Stringari S 1999 {\em Rev. Mod.
  Phys.\/} {\bf 71} 463--512

\bibitem{Pethick2008book}
Pethick C~J and Smith H 2008 {\em {Bose{\textendash}Einstein Condensation in
  Dilute Gases}\/} (Cambridge University Press) ISBN 978-052184651-6

\bibitem{Andrews1997sci}
Andrews M~R, Townsend C~G, Miesner H~J, Durfee D~S, Kurn D~M and Ketterle W
  1997 {\em Science\/} {\bf 275} 637--641

\bibitem{Daniel2013pra}
Daniel A, Agou R, Amit O, Groswasser D, Japha Y and Folman R 2013 {\em Phys.
  Rev. A\/} {\bf 87} 063402

\bibitem{Ketterle1999colle}
W~Ketterle DS~Durfee D~S~K 1999 Making, probing and understanding bose-einstein
  condensates {\em Bose-Einstein Condensation in Atomic Gases, Proceedings of
  the International School of Physics Enrico Fermi, Course CXL\/} ed M~Inguscio
  S~Stringari C~E~W (Amsterdam: IOS Press) pp 67--176

\bibitem{Anderson1995sci}
Anderson M~H, Ensher J~R, Matthews M~R, Wieman C~E and Cornell E~A 1995 {\em
  Science\/} {\bf 269} 198--201

\bibitem{Davis1995prl}
Davis K~B, Mewes M~O, Andrews M~R, van Druten N~J, Durfee D~S, Kurn D~M and
  Ketterle W 1995 {\em Phys. Rev. Lett.\/} {\bf 75} 3969

\bibitem{Bradley1995prl}
Bradley C~C, Sackett C~A, Tollett J~J and Hulet R~G 1995 {\em Phys. Rev.
  Lett.\/} {\bf 75} 1687

\bibitem{deEscobar2008pra}
de~Escobar Y~N~M, Mickelson P~G, Pellegrini P, Nagel S~B, Traverso A, Yan M,
  C{\ifmmode\hat{o}\else\^{o}\fi}t{\ifmmode\acute{e}\else\'{e}\fi} R and
  Killian T~C 2008 {\em Phys. Rev. A\/} {\bf 78} 062708

\bibitem{Stellmer2009prl}
Stellmer S, Tey M~K, Huang B, Grimm R and Schreck F 2009 {\em Phys. Rev.
  Lett.\/} {\bf 103} 200401

\bibitem{Scully1997book}
Scully M~O and Zubairy M~S 1997 {\em {Quantum Optics}\/} (Cambridge University
  Press) ISBN 978-052143595-6

\bibitem{Aspect1988prl}
Aspect A, Arimondo E, Kaiser R, Vansteenkiste N and Cohen-Tannoudji C 1988 {\em
  Phys. Rev. Lett.\/} {\bf 61} 826--829

\bibitem{Aspect1989josab}
Aspect A, Arimondo E, Kaiser R, Vansteenkiste N and Cohen-Tannoudji C 1989 {\em
  J. Opt. Soc. Am. B\/} {\bf 6} 2112--2124

\bibitem{Lin2016jpb}
Lin Y~J and Spielman I~B 2016 {\em J. Phys. B: At. Mol. Opt. Phys.\/} {\bf 49}
  183001

\end{thebibliography}

\end{document}